\documentclass{article}
\usepackage{spconf, amsmath, epsfig, amssymb, booktabs, soul, color, xcolor}

\usepackage{hyperref}
\hypersetup{
    colorlinks=true,
    linkcolor=blue,
    urlcolor=red,
    citecolor=blue}

\title{Dual-Stream Attention Network for Hyperspectral Image Unmixing}

\name{Yufang Wang$^{1,2}$, Wenmin Wu$^{1,2}$, Lin Qi$^{1,2*}$, Feng Gao$^{1,2}$}
\address{$^1$ School of Computer Science and Technology, Ocean University of China, Qingdao 266100, China \\
$^2$ Institute of Marine Development, Ocean University of China, Qingdao 266100, China
\thanks{This work was supported in part by the National Key R\&D Program of China under Grant 2022ZD0117201, and in part by the National Natural Science Foundation of China under Grant 42106191 and Grant U22A2096. (Corresponding author: Lin Qi, Email: qilin2020@ouc.edu.cn)}}

\begin{document}
\maketitle

\begin{abstract}

Hyperspectral image (HSI) contains abundant spatial and spectral information, making it highly valuable for unmixing. In this paper, we propose a Dual-Stream Attention Network (DSANet) for HSI unmixing. The endmembers and abundance of a pixel in HSI have high correlations with its adjacent pixels. Therefore, we adopt a “many to one" strategy to estimate the abundance of the central pixel. In addition, we adopt multiview spectral method, dividing spectral bands into multiple partitions with low correlations to estimate abundances. To aggregate the estimated abundances for complementary from the two branches, we design a cross-fusion attention network to enhance valuable information. Extensive experiments have been conducted on two real datasets, which demonstrate the effectiveness of our DSANet.

\end{abstract}


\section{Introduction}

The restricted spatial resolution of hyperspectral image (HSI), coupled with the rich dimensional information, often results in a high proportion of mixed pixels \cite{c1}. To enhance the application of hyperspectral data in remote sensing,  it is essential to effectively separate mixed pixels, identify endmembers that represent distinct materials, and accurately determine the abundance of each endmember. These steps represent the fundamental objectives in the unmixing process of hyperspectral images \cite{c2, c3}.

With the tremendous achievements of deep learning in other fields, numerous deep learning methods have been introduced to enhance unmixing efficiency. CNNAEU \cite{c8}, SSAE \cite{ssae} and TANet \cite{ta} contain a spatial-spectral autoencoder to reconstruct pixels in a “many to one” strategy. SAWU-Net \cite{sawu} utilizes a spatial attention weighted method in an end-to-end manner to extract spatial features. However, these methods only consider the spatial information and ignore the spectral information.

To make full use of the large continuous spectral bands in HSI, a multiview data generation method is proposed in RMSU \cite{rmsu}, combining the multiview learning and priori information. In SSCU-Net \cite{sscu}, MCSU \cite{mcsu} and MSNet \cite{yangzimo}, spatial-spectral collaborative unmixing is proposed by sharing weights of spatial and spectral feature extraction branches. However, the method of sharing weight can not combine spatial and spectral information sufficiently.

To address the problems mentioned above, we propose a dual-stream attention network which estimates abundance with consideration of both spatial and spectral. In the first branch module, we input a pixel and its adjacent pixels, making full use of the spatial information. In the second branch module, we divide the spectral bands into different partitions to fully utilize the spectral information. Finally, the two estimated abundance are aggregated through a cross-fusion attention network. The experiments on Urban and Jasper Ridge datasets fully demonstrate the superiority of our method.

\section{Methodology}

\begin{figure*}[htb]
\centering
\includegraphics [width=6.5in]{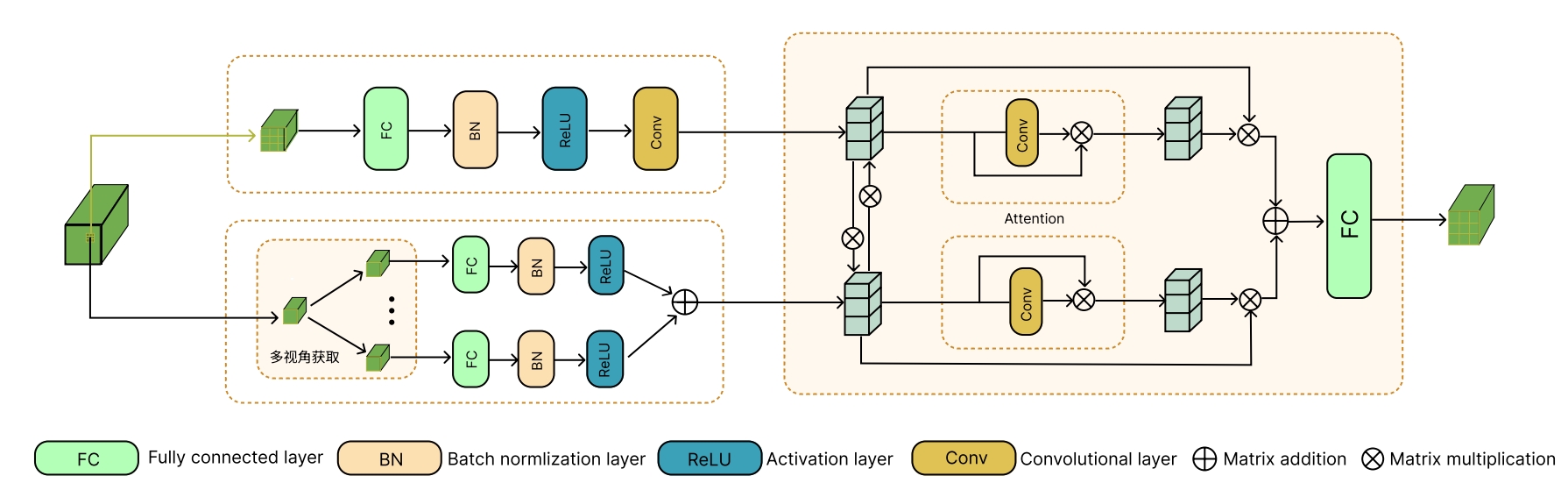}
\caption{Architecture of the proposed DSAnet.}
\label{framwork} 
\end{figure*}

We propose a dual-stream attention network (DSANet), which is composed of two branches and a cross-fusion attention network. The two branches, full-view spatial network and multiview spectral network, are fused through the cross-fusion attention network. The detailed framework of our DSANet is shown in Fig. \ref{framwork}.

\subsection{Full-View Spatial Network}

In hyperspectral image, one matter contained in a pixel likely tends to be contained in its adjacent regions, that is, the endmember appearing in a pixel has a high probability to appear in its adjacent pixels. Based on this, in order to make full use of the spatial information, we utilize the adjacent regions of a pixel for abundance estimation.

As shown in Fig.\ref{framwork}, we select a square window of size $K$, containing $K$ pixels $X\in \mathbb{R}^{K\times L}=\{x_1, x_2, x_3, ..., x_K\}$ with $L$ bands. Then slide the window pixel by pixel to estimate the abundance of its center pixel. During the process, the pixels between in the window need to be encoded to get hidden representation $h_i$: all pixels are passed through a fully connected layer, then processed by a batch normlization layer, and then through a dropout layer, and finally activated by ReLU. The process can be written as follows:
\begin{equation}
\begin{aligned}
   h_i = \textrm{ReLU}(\textrm{Dropout}&(\textrm{BN}(Wx_k)),\\  i = 1, 2, 3,..&., K
\end{aligned}
\end{equation}

Where $W$ represents the weight of the fully connected layer. $H=\{h_1, h_2, h_3, ..., h_K\}$ represents the hidden representations of all pixels in a window. Then we use a convolutional layer to extract features from the hidden representations, which is the estimated endmembers $s_{spa}$ of the center pixel from the full-view spatial network. The formula is expressed as:

\begin{equation}
\begin{aligned}
  s_{spa} = &\textrm{Conv}(H),\\ H=\{h_1, &h_2, h_3, ..., h_K\}
\end{aligned}
\end{equation}

where Conv represents the 1D convolutional layer.

\subsection{Multiview Spectral Network}

Hyperspectral image contains not only rich spatial information, but also multiple spectral information. Therefore, in order to make better use of the spectral information, we adopt a multiview spectral network as supplement to full-view spatial network. The multiview spectral network consists of two modules, the first module is the division of spectral perspectives, and the second module is feature extraction and fusion from each perspective.

Low similarity between every spectral partition is beneficial for endmember extraction. Therefore, in the division of spectral bands, we first divide the bands into $M$ clusters by similarity analysis. There are all bands with high similarity in each cluster. Then we redistribute the bands in clusters approving equal interval sampling between each cluster to obtain $N$ spectral perspectives with low similarity.

In the second module, endmember extraction is performed separately in $N$ perspectives. The estimated endmembers $S_i$ of $N$ groups are fused by matrix addition to generate the final output $s_{spe}$ of the multiview spectral network. The process can be expressed as:

\begin{equation}
\begin{aligned}
   s_i = \textrm{ReLU}(\textrm{Dropout}&(\textrm{BN}(Wx_i)),\\  i = 1, 2, 3,..&., N
\end{aligned}
\end{equation}

\begin{equation}
\begin{aligned}
   s_{spe} = S_1 + s_2 + s_3 +...+s_N,\\  i = 1, 2, 3,..&., N
\end{aligned}
\end{equation}

where $W$ represents the weight of the fully connected layer, BN($\cdot$) represents the batch normlization layer and Dropout($\cdot$) represents the dropout layer.

\subsection{Cross-Fusion Attention Network}

The full-view spatial network makes use of adjacent pixels in HSI fully considering the spatial information, and the multiview spectral network utilizes multiple spectral perspectives to fully capture the spectral information. To take advantage of the complementary information between spatial and spectral, we propose a cross-fusion attention network (CFAN).There are two key parts in our CFAN: weighted cross-fusion attention and endmember extraction. 

\begin{figure*}[htb]
\centering
\includegraphics [width=6in]{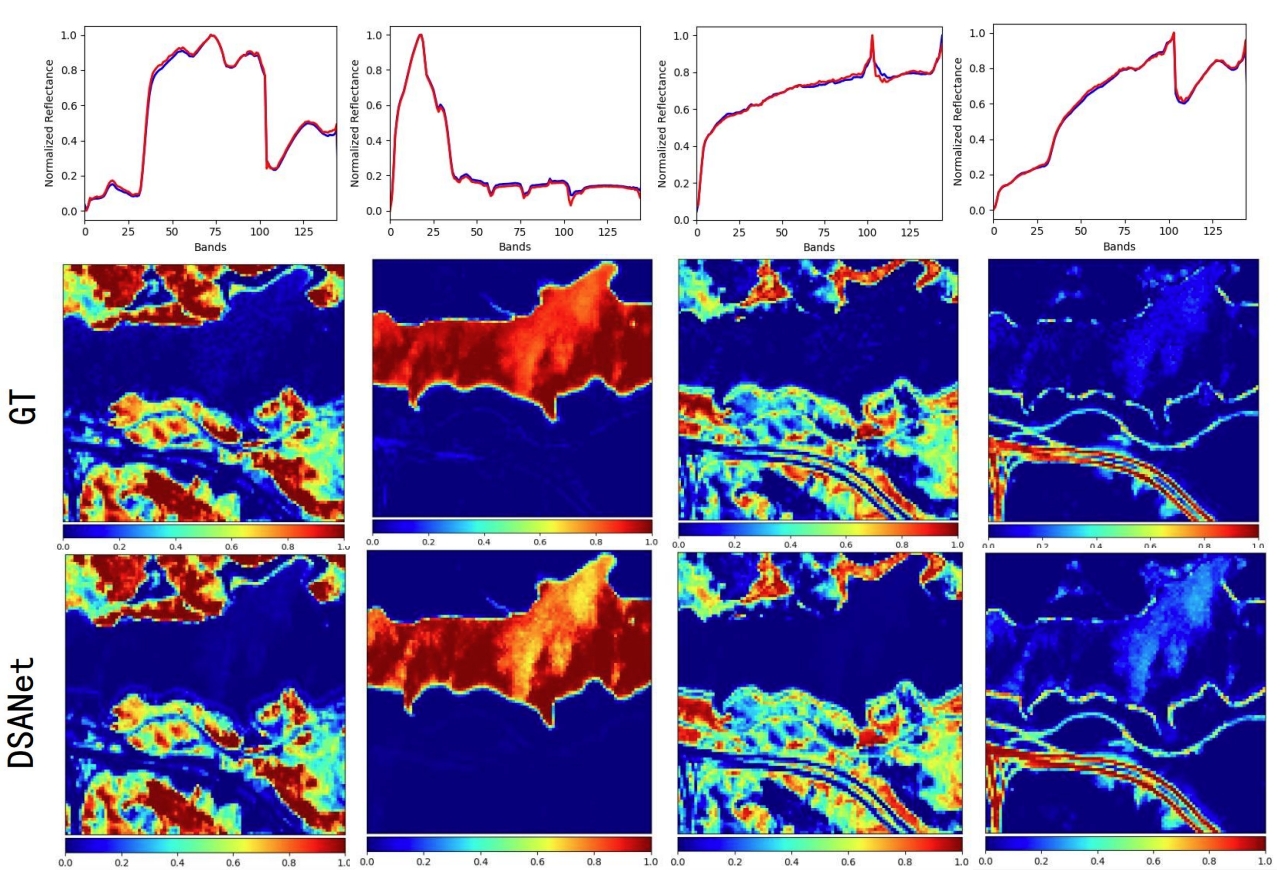}
\caption{Visualization of endmembers (red curves represent groundtruth) and abundance maps (from left to right are Tree, Water, Siol and Road) from Jasper Ridge dataset of our DSAnet.}
\label{experiment} 
\end{figure*}

Firstly, two abundance maps $s_{spa}$ and $s_{spe}$ containing spatial and spectral information are interacted through matrix multiplication to get two interacted abundance maps $s_{spa}'$ and $s_{spa}'$, which highly correlated information is enhanced and low-correlation information is suppressed. Then, a weighted attention operation is carried out on the two interacted abundance maps to enhance self-correlation. Than the two interacted abundance maps are weighted by their initial abundance maps respectively. Finally the aggregated abundance map $s_c$ of the center pixel are generated. The operation can be expressed as follows:

\begin{equation}
    s'_{spa} = s_{spa} \cdot s_{spe}, 
\end{equation}

\begin{equation}
    s'_{spe} = s_{spe} \cdot s_{spa}, 
\end{equation}

\begin{equation}
    s = s_{pa}\cdot att(s_{spa}') + s_{spe}\cdot att(s_{spe}')
\end{equation}

where $att(\cdot)$ denotes attention layer. Secondly, a fully connected layer is used to extract endmembers, which weight represents the extracted endmembers. The process is expressed as follows:

\begin{equation}
    \hat{x} = W\cdot s_c
\end{equation}

where $W$ represents the weight of the fully connected layer. The loss function of our DSANet is defined as follows:

\begin{equation}
    L(i) = \lambda_1 cos^{-1} \frac{\langle x_{i,j}, \hat{x}_{i,j} \rangle }{|x_{i,j}|_2 |\hat{x}_{i,j}|_2}+\lambda _2 |s_c|_{\frac{1}{2}}
\end{equation}

Where $x_{i,j}$ is the original center pixel in HSI, and $\hat{x}_{i,j}$ is the reconstructed center pixel, and $s_c$ represents the abundance vector of the center pixel.

\begin{table}[ht]
\centering
\caption{SAD($\times10^{-2}$) for the Urban dataset.}
\vspace{0.5em}
\setlength{\tabcolsep}{1.1mm}
\small
\begin{tabular}{c|c c c c c}
\toprule
EM & CNNAEU & EndNet & SSAE & MSNet & Ours  \\
\midrule
Asphalt & 5.75$\pm$0.5 & 5.98$\pm$0.2 & 6.02$\pm$0.5 & \textbf{4.38$\pm$0.3} &  4.40 $\pm$ 0.2 \\ 
Grass & 3.66$\pm$0.4 & 5.34$\pm$0.1 & 3.59$\pm$0.3 & 3.28$\pm$0.2 &  \textbf{3.10$\pm$0.2} \\ 
Tree & 3.21$\pm$0.3 & 4.57$\pm$0.2 & 3.18$\pm$0.2 & 3.27$\pm$0.3 & \textbf{3.15$\pm$0.2}  \\
Roof & 3.32$\pm$0.6 & 3.89$\pm$0.7 & 2.76$\pm$0.3 &  2.49$\pm$0.4 & \textbf{2.37$\pm$0.3} \\ 
\midrule
Average & 3.98$\pm$0.3 & 4.95$\pm$0.3 & 3.89$\pm$0.1 & 3.35$\pm$0.1 & \textbf{3.25$\pm$0.1}\\ 
\bottomrule
\end{tabular}
\label{table1}
\end{table}

\begin{table}[ht]
\centering
\caption{RMSE($\times10^{-2}$) for the Urban dataset.}
\vspace{0.5em}
\setlength{\tabcolsep}{1.1mm}
\small
\begin{tabular}{c|c c c c c}
\toprule
EM & CNNAEU & EndNet & SSAE & MSNet & Ours  \\
\midrule
Asphalt & 12.49$\pm$0.3 & 10.62$\pm$0.1 & 9.75$\pm$0.4 & \textbf{8.83$\pm$0.2} &  8.92$\pm$0.3 \\ 
Grass & 12.57$\pm$0.3 & 13.85$\pm$0.3 & 9.59$\pm$0.2 & 7.92$\pm$0.2 &  \textbf{7.41$\pm$0.2} \\ 
Tree & 8.55$\pm$0.2 & 9.07$\pm$0.3 & 5.87$\pm$0.2 & 5.44$\pm$0.2 & \textbf{5.32$\pm$0.2}  \\
Roof & 8.54$\pm$0.2 & 6.51$\pm$0.2 & 5.76$\pm$0.2 &  7.34$\pm$0.3 & \textbf{7.27$\pm$0.2} \\ 
\midrule
Average & 10.54$\pm$0.1 & 10.02$\pm$0.2 & 7.75$\pm$0.1 & 7.38$\pm$0.1 & \textbf{7.23$\pm$0.2} \\ 
\bottomrule
\end{tabular}
\label{table2}
\end{table}

\begin{table}[ht]
\centering
\caption{SAD($\times10^{-2}$) for the Jasper Ridge dataset.}
\vspace{0.5em}
\setlength{\tabcolsep}{1.1mm}
\small
\begin{tabular}{c|c c c c c}
\toprule
EM & CNNAEU & EndNet & SSAE & MSNet & Ours  \\
\midrule
Asphalt & 11.94$\pm$2.1 & 4.57$\pm$0.4 & 3.37$\pm$0.1 & 3.05$\pm$0.1 &  \textbf{2.05$\pm$0.2} \\ 
Grass & 6.92$\pm$0.4 & 5.05$\pm$0.9 & 4.72$\pm$0.1 & \textbf{3.33$\pm$0.1} &  3.64$\pm$0.1 \\ 
Tree & 10.15$\pm$0.9 & 5.29$\pm$0.3 & 3.02$\pm$0.3 & 2.33$\pm$0.3 & \textbf{1.65$\pm$0.2}  \\
Roof & 7.45$\pm$0.4 & 3.54$\pm$0.2 & 2.77$\pm$0.2 &  2.07$\pm$0.4 & \textbf{2.05$\pm$0.2} \\ 
\midrule
Average & 9.12$\pm$0.6 & 4.61$\pm$0.5 & 3.47$\pm$0.1 & 2.70$\pm$0.0 & \textbf{2.34$\pm$0.2}\\ 
\bottomrule
\end{tabular}
\label{table3}
\end{table}

\begin{table}[ht]
\centering
\caption{RMSE($\times10^{-2}$) for the Jasper Ridge dataset.}
\vspace{0.5em}
\setlength{\tabcolsep}{1.1mm}
\small
\begin{tabular}{c|c c c c c}
\toprule
EM & CNNAEU & EndNet & SSAE & MSNet & Ours  \\
\midrule
Asphalt & 13.56$\pm$0.3 & 8.85$\pm$0.4 & 5.15$\pm$0.6 & 3.73$\pm$0.7 &  \textbf{3.12$\pm$0.3} \\ 
Grass & 9.66$\pm$0.3 & 6.88$\pm$0.3 & 5.63$\pm$0.4 & \textbf{2.86$\pm$0.5} &  3.01$\pm$0.3 \\ 
Tree & 10.61$\pm$0.2 & 10.59$\pm$0.2 & 6.19$\pm$0.4 & 4.95$\pm$0.6 & \textbf{4.08$\pm$0.2}  \\
Roof & 8.64$\pm$0.2 & 11.17$\pm$0.4 & 7.15$\pm$0.3 &  3.91$\pm$0.3 & \textbf{3.87$\pm$0.2} \\ 
\midrule
Average & 10.62$\pm$0.1 & 9.37$\pm$0.5 & 6.03$\pm$0.2 & 3.86$\pm$0.3 & \textbf{3.52$\pm$0.3} \\ 
\bottomrule
\end{tabular}
\label{table4}
\end{table}

\section{Experimental Results and Analysis}

The datassets used in our experiments are Urban\cite{urban} and Jasper Ridge\cite{jasper} datasets. Urban dataset contains 162 spectral bands which ground truth contains 4 endmembers: tree, water, soil and road. Jasper Ridge dataset contains 198 spectral bands which ground truth contains 4 endmembers: asphalt, grass, tree, and roof. To evaluate the effectiveness of our proposed network, spectral angle distance $SAD = cos^{-1} \frac{\langle e, \hat{e} \rangle}{|e|_2 |\hat{e}|_2}$ and root mean square error $RMSE = \sqrt{\frac{1}{N} \sum\limits^N_{i=1} (s_i - \hat{s_i} )^2 }$ are utilized.

CNNAEU\cite{c8}, EndNe\cite{c11}, SSAE\cite{ssae} and MSNet\cite{yangzimo}  are used for comparative analysis with our DSANet. Table \ref{table1} and Table \ref{table2} show the performance of each method on the Urban dataset, and Table \ref{table3} and Table \ref{table4} show the performance of each method on the Jasper Ridge dataset. It can be seen that our DSANet can capture more accurate endmembers and batter abundance. Fig.\ref{experiment} shows the visualization of endmembers and abundance maps of our DSANet and groundtruth.

\section{Conclusion}
In order to fully utilize the spatial and spectral information in HSI, we propose a dual-stream attention network for hyperspectral unmixing. The network has two branch networks and a cross fusion attention network. The first branch full-view spatial network fullly utilizes spatial information from adjacent pixels. The second branch multiview spectral network divides spectral bands into multiple views, fully utilizing the diversity and multiplicity of spectral bands. The  cross fusion attention aggregated the two branches and extract endmembers. The experiment results in two real datasets indicate that our proposed method performs better than other state-of-the-art methods.

\end{document}